\documentclass[twocolumn,amsmath,amssymb,prl]{revtex4}
\usepackage{graphicx}
\usepackage{dcolumn}
\usepackage{bm}

\def\mnras{MNRAS}

\def\beq{\begin{equation}}
\def\eeq{\end{equation}}

\begin{document}
\input{epsf}

\title{Eavesdropping on Radio Broadcasts from Galactic Civilizations with
Upcoming Observatories for Redshifted 21cm Radiation}

\author{Abraham Loeb \& Matias Zaldarriaga}
\affiliation{Astronomy Department, Harvard University, 60 Garden Street,
Cambridge, MA 02138}

\begin{abstract}

The question of whether intelligent life exists elsewhere is one of the
fundamental unknowns about our Universe.  Over the past decade $\gtrsim
200$ extra-solar planets have been discovered, providing new urgency for
addressing this question in these or other planetary systems.
Independently of this perspective, new radio observatories for cosmology
are currently being constructed with the goal of detecting 21cm emission
from cosmic hydrogen in the redshift range $6\lesssim z\lesssim 15$. The
radio frequency band covered by these experiments overlaps with the range
of frequencies used for telecommunication on Earth, a regime that was never
explored with high sensitivity before.  For example, the Low-Frequency
Demonstrator (LFD) of the Mileura Wide-Field Array (MWA), will cover in 8
kHz bins the entire frequency range of 80--300 MHz, which is perfectly
matched to the band over which our civilization emits most of its radio
power.  We show that this and other low-frequency observatories
(culminating with the Square Kilometer Array [SKA]) will be able to detect
radio broadcast leakage from an Earth-like civilization out to a distance
of $\sim 10^{1-2.7}$pc, within a spherical volume containing
$10^{(3-8)}\times (\Omega_b/4\pi)^\alpha$ stars, where $\alpha=1$ (or 1.5)
for a radar beam of solid angle $\Omega_b$ that remains steady (or sweeps)
across the sky. Such a radio signal will show-up as a series of narrow
spectral lines that do not coincide with known atomic or molecular
lines. The high spectral resolution attainable with the upcoming
observatories will allow to monitor the periodic Doppler shift of the
broadcasted lines over the planet's orbital period around the parent star.
Determination of the parent star mass through observations of its spectrum
could then be used to infer the inclination, semi-major axis and
eccentricity of the planet's orbit. This, in turn, will allow to estimate
the temperature on the planet's surface and to assess whether it can
support liquid water or life as we know it.

\end{abstract}


\maketitle

\section{I. Introduction}

Over the past decade, more than 200 new planets outside the solar system
were discovered \cite{Marcy}. Because the majority of these planets were
detected through precise radial velocity measurements, most of the known
extra-solar planets have a mass comparable to that of Jupiter and a tight
orbit around a star within a hundred parsec from the sun.  Habitable
conditions for life may exist on moons around these giant planets or in
other planetary systems that better resemble the solar system.  Given the
high detection rate of new planets, the question of whether intelligent
life exists elsewhere in the Galaxy is particularly timely.

The search for extra-terrestrial intelligence (SETI) has a long history
\cite{Wilson}.  In 1960, a targeted search of two nearby sun-like stars
used the 25-meter radio telescope of NRAO \cite{seven}. This project OZMA
\cite{eight} was followed by other initiatives using radio telescopes
\cite{nine} and some SETI programs continue today (http://www.seti.org/).
NASA's program {\it The High Resolution Microwave Search} (HRMS) included
both a targeted search at individual stars as well as an all sky survey,
but was cancelled by Congress in the early 1990s \cite{ten}.  Funding for
SETI initiatives today originates from non-profit organizations such as the
Planetary Society \cite{eleven} and the SETI Institute \cite{twelve}. The
SETI Institute has revived the targeted search component of HRMS through
project PHOENIX. The SERENDIP program at UC Berkeley, as well as the joint
venture between the SETI Institute and UC Berkeley involving the
construction of the {\it Allen Telescope Array}
(http://astron.berkeley.edu/ral/ata/), perform SETI work on radio
telescopes that are also used for astronomical studies.

Most of the radio emission from our civilization on Earth originates in the
frequency range of 50--400MHz. Table I summarizes the main sources of radio
emission, including telecommunication and military radars.  The bandwidth
of the signals is narrow, with a fractional frequency width of
$(\delta\nu/\nu)_{\rm int} \sim 10^{-9}$ for TV and FM stations and
$10^{-5}$ for the military radars.  From a distance, the broadcasted power as
a function of frequency would appear as narrow spectral lines.

For practical purposes, the traditional SETI searches focused on high radio
frequencies that do not match the frequency range over which our
civilization is most luminous (see Table I).  Coincidentally, a new
generation of radio observatories is being constructed around the globe
with the primary goal of probing the redshifted 21cm (1420 MHz) emission
from neutral hydrogen before it was ionized by the first galaxies. As these
observatories focus on the redshift range of $z=6$--15, they cover the
frequency range of 80--300 MHz. This frequency range happens to coincide
with the band over which our civilization is most luminous.
Coincidentally, these sensitive new instruments could be suited for SETI
work in this frequency band. The goal of this paper is to characterize and
motivate the use of these new observatories for SETI studies.  SETI work
requires frequency resolution which happens to coincide with the design for
one of these forthcoming observatories, the {Mileura Wide-Field Array}
(MWA).  We therefore use this experiment as a benchmark example to
illustrate our main conclusions.

\section{II. 21 cm Cosmology}

\subsection{II.1 Background}
 

The targeted cosmic fluctuations in the 21 cm brightness temperature are
small, around 10--20 mK, depending on redshift and angular scale.
Detection of these fluctuations is particularly challenging since the sky
brightness temperature at the same wavelength is four orders of magnitude
higher \cite{Loeb,FBO}. The faintness of the signal will drive the design
of instruments to have large collecting areas. During the cosmic epoch of
reionization, when the universe is partially ionized, the 21 cm signal has
a ``swiss cheese" topology created by the bubbles of ionized hydrogen
(which are invisible at 21cm) surrounded by regions in which hydrogen is
neutral and can thus be observed \cite{Loeb}. The characteristic bubble
size grows during reionization and eventually reaches the large scale of
tens of comoving Mpc at the end of reionization, depending on the detailed
characteristics of the sources of radiation and the distribution of the gas
\cite{WL,FZH,Zahn,Shapiro,furl-21cmsim,sok01,sh03a}. Getting a good measure
of the size distribution of these bubbles and other statistical properties
of the fluctuations will require mapping tens to hundreds of square degrees
(necessary in order to probe many large bubbles at the end of reionization)
with at least arcminute resolution (in order to capture the many small
bubbles present at the beginning of reionization).  The removal of the
strong synchrotron foreground will be achieved by subtracting the sky maps
at slightly different frequencies \cite{FBO}. For the cosmological signal,
this is equivalent to slicing the hydrogen distribution at different
redshifts and hence one should see a different map of its bubble structure,
while the synchrotron foreground remains nearly the same.

Given that 21cm cosmology is a new well-funded frontier \cite{Loeb,FBO}, a
promising strategy for SETI would be to piggyback on the radio
observatories under construction. Next we summarize the characteristics of
MWA and other forthcoming experiments and consider their sensitivity for
SETI work.

\subsection{II.2 Upcoming Low-Frequency Observatories}

Several observational groups around the globe are constructing
low-frequency arrays that will be capable of mapping the three-dimensional
distribution of cosmic hydrogen in the infant universe. These arrays are
aiming to detect the long-wavelength (redshifted 21cm) radio emission from
hydrogen atoms, and map the three-dimensional distribution of cosmic
hydrogen by making measurements over a wide range of frequencies.

The Low-Frequency Demonstrator (LFD) of the Mileura Wide-Field Array (MWA;
see http://web.haystack.mit.edu/arrays/MWA/ for details) is currently under
construction, with completion expected in early 2008. The MWA-LFD will
cover the frequency range of 80--300MHz in 8kHz bins and include 500
dipole-based antenna tiles. Each 4m$\times$4m tile will contain 16 dipole
antennas.  Altogether the total collecting area will be 8000 m$^2$ at 150
MHz. The tiles will be scattered across a 1.5 km region, providing an
angular resolution of a few arcminutes. MWA will be located in a
radio-quiet site in Western Australia, called Mileura.  Other experiments
whose goal is to detect 21cm fluctuations from the epoch of reionization at
$z\sim 6-12$ include the Low-Frequency Array ({\it LOFAR}; {\it
http://www.lofar.org}), the Primeval Structure Telescope ({\it PAST}; {\it
http://arxiv.org/abs/astro-ph/0502029}),
and in the more distant future expansions of MWA-LFD with the ultimate
observatory being the Square Kilometer Array ({\it SKA}; {\it
http://www.skatelescope.org}). 


\section{III. Signal Detectability for SETI}

It is difficult to gauge the likelihood of success for the SETI endeavor
\cite{Marko}.  As a result, ambitious and thus expensive projects whose
sole purpose is SETI are unlikely to be funded. It is thus particularly
fruitful to understand the synergies and complementarities of other
scientific projects with various types of SETI searches.

To motivate the potential of this approach we first summarize the expected
sensitivity of MWA-LFD and other experiments. To assess what such detection
levels could mean for SETI we use the current radio emission by our
civilization as an example and also compare with other past and future SETI
studies.

It is important to note that high-redshift 21 cm surveys will operate at
different frequencies than past and current SETI searches, making them
complementary. They will also have very different observing strategies and
so they will explore a different part of the signal parameter space.

\subsection{III.1 Possible Signals: Our Civilization as an Example}

Forecasting how the signals from another civilization might look like is
not an easy proposition. It is clear from the start that although one might
be able to come up with educated guesses it is very possible that if we
ever detect such a signal it will look different from our initial guesses.

For example, one might argue that the characteristics of the signal might
be very different if the signal we receive was broadcasted intentionally by
another civilization to announce its presence, a ``beacon" of sorts, or
just the leakage of signals that the extra-terrestrials produce for their
own communication either within their own planet or between
spacecrafts \footnote{Note that aside from planet-based broadcasting, there
is a possibility of discovering communication stations that were
distributed throughout the Galaxy by another civilization which by itself
resides far away (G. Laughlin, private communication).}.

Whether we look for a beacon or unintentional leakage, the signal we are
after must posses some characteristic that will enable us to distinguish it
from naturally produced signals. Examples of such characteristics might be
peculiar time variability and very narrow frequency structure centered on a
frequency where there are no atomic or molecular lines.

A good starting point when considering ``leakage"-type signals is the
signal from our own civilization. A summary of the strongest emission
component from the Earth is given in Table \ref{earth} (adopted from
Ref. \cite{sullivan}). The strongest sources of emission in terms of their
power are military radars \footnote{Table I indicates that it would be
easier to detect militant civilizations than their peaceful analogs. If
this extrapolation from our own experience is to be believed, then we
should make any effort to limit communication with the brightest
civilizations in a survey.} ($[\delta\nu/\nu]_{\rm int}\sim 10^{-5}$),
followed by TV stations and FM radio stations ($[\delta\nu/\nu]_{\rm int}
\sim 10^{-9}$). In terms of power per Hz, the strongest signal originates
from TV stations.  These two signals are emitted in a fairly isotropic
way. If our civilization was using a telescope to beam those signals as a
beacon or perhaps to communicate with a distant spacecraft, then the
effective emitted power could have been much larger. Of course in that case
only a correspondingly smaller fraction of distant observers would detect
the signal at any given time. It is possible then that the signal we are
after, would be much stronger than our own emission but with a small duty
cycle.

It is important to keep in mind that various physical processes would
broaden the bandwidth or modulate the frequency of the emitted radio
signal. In particular the motion of the planet around its parent star and
the spin of the planet, both introduce shifts. The shifts are
especially important for long integration times with narrow bandwidths. For
the case of the Earth, the spin velocity $v_{\rm spin}$ would introduce a
shift of order,
\begin{eqnarray}
\left({\delta \nu \over \nu}\right)_{\rm spin}& =& {v_{\rm spin} \over c}
\approx 1.5 \times \ 10^{-6} ; \\ \delta \nu_{\rm spin} & \approx & 0.15\
{\rm kHz} \times \left( {\nu \over 100 \ {\rm MHz}}\right);
\end{eqnarray}
over the course of a day.  The envelope of this variability will be
modulated by the orbital velocity $v_{\rm orb}$ around the sun over the
course of a year with,
\begin{eqnarray}
\left({\delta \nu \over \nu}\right)_{\rm orb}& =& {v_{\rm orb} \over c}
\approx 10^{-4} ; \\ \delta \nu_{\rm orb} & \approx & 10\ {\rm kHz} \times
\left( {\nu \over 100 \ {\rm MHz}}\right).
\end{eqnarray}
The significance of these frequency drifts will depend on the duration of
the observations which could be very long, of order months, in the context
of 21 cm experiments.

If the transmitters are not distributed uniformly on the surface of the
planet, the planet's spin will also introduce a modulation in the intensity
of the signal. Of course if the signal originates from a beacon then it
might be significantly beamed or pulsed. The modulations of both intensity
and frequency could eventually become clues that the emission originates
from another civilization or at least from a planet around a star.

In summary, the generic property of the expected signals are that they are
narrow in frequency and that they might have interesting time variability
as a result of both the orbital motion of the planet and its rotation
around its axis, in addition to the variability introduced at the
transmission.

Other observing techniques should be used to follow-up on any radio
detection of planets in an attempt to check whether it could host life as
we know it. The {\it habitable zone} is the range of orbital radii over
which liquid water may exist on a rocky planet, so that life as we know it
could develop there based on a similar network of chemical reactions to the
one that existed on the early Earth. Obviously, the habitable zone
straggles the distance of 1AU for a sun-like star, but its location changes
for other stellar types \cite{Lopez,Forget}. The first extensive discussion
on habitability was provided by Kasting et al. \cite{Kasting,Kasting2}.
Existing observational techniques could find planets in the habitable zone
in the future \cite{Charb}. Proposed space missions, such as the
Terrestrial Planet Finder
(http://tpf.jpl.nasa.gov/earthlike/earth-like.html) and Darwin
(http://sci.esa.int/science-e/www/area/index.cfm?fareaid=28), focus on
searches for signatures of life on extrasolar planets.  

So far, no extrasolar planet was found to reside within the habitable zone
of its parent star (but see the indirect constraints in Refs.
\cite{Jones,Raymond}). Obviously, all planets which will be found to reside
in the habitable zone should be searched with great sensitivity for radio
signals, and vice versa.  The different search methods are complementary
and one should use all probes at our disposal to constrain the physical
conditions on such planets and determine whether they might support
life. This obviously includes spectroscopy of the parent star, which would
yield its distance, spectral type, as well as the location of its habitable
zone.

\subsection{III.2 Experiment Characteristics}

The 21 cm signal from high-redshift hydrogen is faint, leading to
brightness temperature fluctuations of the order of $20$ mK on top of a sky
brightness of $\sim 200$ K dominated by Galactic synchrotron radiation
\cite{FBO}.  The HII regions responsible for a large part of the signal
subtend tens of arcminutes on the sky \cite{WL}, and so large fields of
view will be required to eliminate uncertainties due to cosmic variance.
The MWA-LFD and other future surveys will be designed to have large
collecting areas to be able to detect the weak 21cm signal, with fields of
view that cover many square degrees on the sky. Observing the 21 cm signal
will also require integrating on those fields for long periods of time.
Finally the technique that will be used to distinguish the high-redshift 21
cm signal from various astrophysical foregrounds and radio frequency
interference involves mapping the signal with high frequency resolution.

As a result, {\it future experiments will have extremely good sensitivity
and could thus place strong limits on potential SETI signals}. The
extensive time coverage would allow the study of signals which might only
be transmitted in our direction for a small fraction of the time.

The point source sensitivity of an interferometer composed of antennae of
individual area $A_a$ and $N_b$ baselines that observes for a time $t_o$
with a bandwidth $\Delta \nu$ is, \beq F_{\rm PSS} = C {2 k T_{sys} \over
A_a \sqrt{\Delta\nu t_o N_b}},
\label{sens}
\eeq where $C$ is a constant of order unity that depends on the spatial
distribution of baselines. The system temperature, $T_{sys}$, is dominated
by the synchrotron emission from our Galaxy and is thus frequency dependent
and variable across the sky. 

For the MWA-LFD, the point source sensitivity at $200$ MHz is\footnote{The
value of $F_{\rm PSS}$ for other frequencies can be found at
http://web.haystack.mit.edu/arrays/MWA/LFD .}, \beq F_{\rm PSS} = 0.4 \
{\rm mJy} \ \left({\Delta \nu \over 8 \ {\rm kHz}}{ t_o \over {\rm
month}}\right)^{-1/2}, \eeq where the current design aims at
$\Delta\nu=8$kHz.  For a given distance to the source $d$, this formula can
be turned into the minimum power required of the transmitter for MWA to see
it,
\begin{eqnarray} P_{\rm min} &=& 3.5 \times 10^{12} {\rm W} \nonumber \\
& \times &
\left({d \over 100 {\rm pc}}\right)^2 \times \left({\Delta \nu \over 8 \
{\rm kHz}}\right)^{1/2} \times \left({t_o \over {\rm month}}\right)^{-1/2}.
\label{pm}
\end{eqnarray}
Here it is assumed that the intrinsic bandwidth of the signal
$\delta\nu_{\rm int}$ is much smaller than $\Delta \nu$. In this regime one
gains by making $\Delta \nu$ smaller.

Note that {\it the power level in equation (\ref{pm}) is only an order of
magnitude larger than the present-day military radar emission from the
Earth (Table I) even for a source which is 100 pc away}. It is plausible
that the actual emission from other civilizations would be stronger. It is
also possible that the emission would be beamed or pulsed so that it would
be visible to us for only a small fraction of the time. The required
minimum power scales as $d^2$. Thus a duty cycle of
$10^{-2}$ requires a source to be merely 10 times closer for the same
emitted peak power.  One way to increase the sensitivity is to reduce the
bandwidth of the observations down to $(\delta \nu)_{\rm int}$ below which
there is no additional gain.

\begin{figure} [ht]
\centerline{\epsfxsize=3.4in \epsfbox{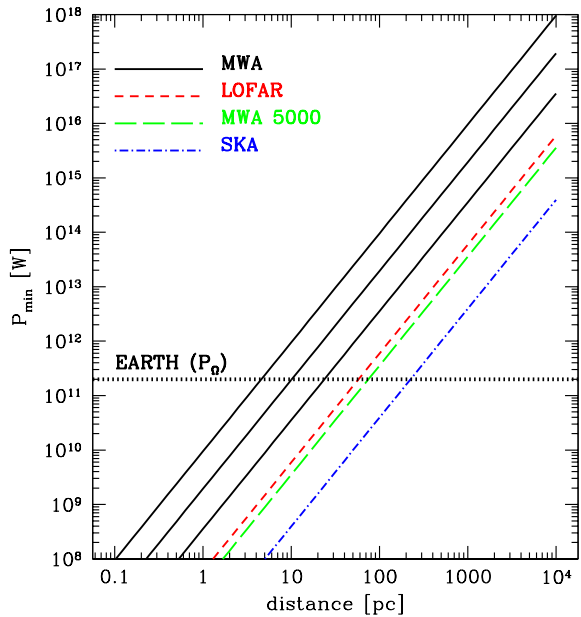}}
\caption{The minimum detectable radio power $P_{\rm min}$ for various
high-redshift 21 cm surveys and observing times as a function of the
distance to the source. We assume for simplicity that the source emits
isotropically and steadily. For the MWA we adopt a bandwidth of
$\Delta\nu=8$ kHz and observing times of 1 hour, 1 day and one month (solid
lines from top to bottom). We assumed the same bandwidth for LOFAR, for a
future extension of MWA with ten times its collecting area (called MWA
5000) and for the SKA.  In those cases we only plot the sensitivity for a
one month integration. The dashed line delineates the power per solid
angle, $P_\Omega$, along the beam of military radars [Ballistic Missile
Early Warning Systems ({\it BMEWS})] from our civilization.}
\label{distance}
\end{figure}

Figure \ref{distance} shows the power required for a transmitter (assuming
isotropic emission) to be detectable by MWA-LFD as a function of distance
to the source. The three solid lines show results for an hour, a day or a
month of integration on a field. For a month of integration time, the Earth
radio signal could be detected out to a distance of $\sim 23$ pc interior
to which there are $\sim 10^{3-4}$ stars over the whole sky (the field of
view of MWA is about a percent of the sky). However, only a fraction
$(\Omega_b/4\pi)^\alpha$ of the military radars may be beamed in our
direction, where $\alpha=1$ (or 1.5) for a radar beam of solid angle
$\Omega_b$ that remains steady (or sweeps) across the sky. We also plot the
expected one-month sensitivity of LOFAR, MWA 5000 (a future extension of
MWA with ten times its collecting area) and the SKA -- which would be able
to detect the Earth out to a distance of 0.2 kpc. We note that the use of
the SKA in the broader context of the exploration of life was discussed in
Ref.  \cite{Lazio}.

Another figure of merit that can be used to compare MWA with other previous
and future SETI searches is the flux, in ${\rm W~m^{-2}}$, that must be
arriving on Earth for a source to be detectable, \beq f_{\rm PSS} = 3
\times 10^{-26} {\rm W}\ {\rm m}^{-2} \times \left({\Delta \nu \over 8\
{\rm kHz}}\right)^{1/2} \times \left({t_o \over {\rm month}}\right)^{-1/2}.
\eeq

The promise of future 21 cm surveys is evident from Figure
\ref{distance}. As the collecting area increases, the arrays will probe
larger distances.

\subsection{III.3 Comparison with Other SETI Probes}

Next we compare the sensitivity of MWA with other SETI probes in the radio
frequency range \cite{tarter}. A very important difference between MWA and
previous probes involves its sensitivity to lower frequencies. Most radio
SETI programs have searched above 1 GHz and around the unredshifted 21 cm
line at 1.42 GHz. There were several reasons to focus on this frequency
range. First, the sky brightness temperature is smaller above 1 GHz as the
synchrotron emission from relativistic electrons in our galaxy becomes
sub-dominant relative to the Cosmic Microwave Background. Second, when
looking for a ``beacon" type signal it could be argued that other
civilizations might center their transmitters around the 1.42 GHz line with
the hope of reaching astronomers using that line for astronomical
observations. It is difficult to gauge whether such an argument carries
much weight.

Of course, our own radio broadcasting is far greater at lower frequencies,
so at least for the purpose of ``eavesdropping" on another civilization,
lower frequencies might be more interesting. The fact that our civilization
makes much use of the lower frequency spectrum presents severe technical
difficulties for SETI programs trying to operate in this frequency range as
they have to filter-out our own radio-frequency interference (RFI). Thus,
21cm cosmology is a case in which an unrelated science driver will open a
new and potentially {\it more suitable} window for SETI programs. The
interest in high-redshift 21 cm surveys means that there will be
significant efforts to control RFI by, for example, placing the
observatories in remote locations with the lowest RFI record (such as
China, Australia, Africa, or even the moon), as well as developing new
filtering techniques for RFI and ionospheric noise. Obviously SETI programs
could benefit significantly from these technological developments.

A good summary of previous and planned SETI programs can be found in
Ref. \cite{tarter}. Here we will only comment on some of the most sensitive
of those programs and focus our attention only on radio band examples. In
doing so, we would like to demonstrate that SETI work with the MWA-LFD and
other 21 cm surveys could have very competitive sensitivities.

The SETI programs carried out at Arecibo and the Parkes telescope in
Australia achieved the best sensitivities with their very large collecting
areas. For example, the Phoenix project carried out by the SETI institute
reached a flux limit of $10^{-26}~{\rm W~m^{-2}}$ by searching the nearest
600 stars at frequencies in the range 1.2--3 GHz with a frequency
resolution of 1 Hz. The SERENDIP IV project out of UC Berkeley also uses
Arecibo to monitor the sky visible from Arecibo up to a flux of $5 \times
10^{-24}~{\rm W~m^{-2}}$. It covers the sky every three years and looks for
recurring signals. It explores the frequency range $1.420\pm 0.05$ GHz with
a 0.6 Hz bandwidth. A companion SERENDIP search with similar
characteristics operates from the Parkes telescope.

The collecting area of Arecibo is approximately $7\times 10^4$ m$^2$,
roughly ten times that of MWA-LFD. By working at higher frequencies where
the sky is less bright, the sensitivity of previous surveys was
enhanced. The MWA-LFD with its deep observations of certain fields will be
competitive or even better than those previous surveys at least for steady
signals.  Production of detailed images from the epoch of reionization will
require an increase in collecting area of future arrays by a factor of
10--100. Arrays of dipoles such as the MWA-LFD, have a much larger field of
view, around 400 square degrees at 200 MHz, while a telescope like Arecibo
has a field of view of less that a tenth of a square degree.  The long
observing times required for high-redshift 21 cm observations also open a
new regime for SETI observations.

The next generation of SETI work will be carried out by the Allen Telescope
Array (ATA), a joint venture of the SETI Institute and UC Berkeley. The
will include about three hundred and fifty 6 m dishes. The plan is to
survey of order $10^5$ stars three times in the frequency range of $1-10$
GHz over a period of 10 years.  This instrument has a collecting area
similar to that of the MWA-LFD but a smaller field of view.  It is
complementary to what could be done with proposed 21 cm arrays over a
different range of frequencies. Again, the higher frequencies lead to a
smaller sky temperature and thus to a better raw sensitivity in terms of
$P_{\rm min}$.  But the larger collecting area of future extensions of the
MWA-LFD will partially compensate for their larger system temperature.  The
observing strategy of high-redshift 21 cm surveys, namely deep integration
on selected fields, is also very different from the ATA survey. It will
open a window to search for other types of signals, such as those that
could be intense but with a short duty cycle, perhaps appearing for a short
time once every rotation period of the planet.

It is very important to stress that all these different SETI projects have
fairly different characteristics in terms of frequency coverage, time spent
observing any given source, or bandwidth of the observation. Given our
ignorance about the characteristics of the extra-terrestrial signal, it is
certainly worth using all these different instruments for the search.  In
this context, it is clear that the redshifted 21 cm observatories could
attain unique sensitivities that were never realized before.

\section{IV. Discussion}

Figure 1 indicates that integration times of months to years on upcoming
low-frequency observatories will provide an unprecedented sensitivity to
radio broadcasts from an Earth-like civilization at Galactic distances of
0.01--0.5 kpc, within a spherical volume containing $10^{(3-8)}\times
(\Omega_b/4\pi)^\alpha$ stars, where $\alpha=1$ (or 1.5) for a radar beam
of solid angle $\Omega_b$ that remains steady (or sweeps) across the sky.
The challenge to detect redshifted 21cm radiation from the infant Universe
will open a new opportunity for eavesdropping on the leakage of signals
from Galactic civilizations in the same band of radio frequencies that our
civilization uses for communication. Such an opportunity never existed
before.

SETI work requires an optimized algorithm and software for extracting the
artificial signal out of the data stream of upcoming observatories aiming
to detect the cosmological 21cm brightness fluctuations. If not searched
for, the desired SETI signal may be lost.  Processing of the data will be
done by filtering-out unwanted foregrounds. In particular, the
radio-communication signal from a faint Earth-like source can be separated
from the frequency structure of the redshifted 21cm fluctuations through
its time dependence.

The desired radio signal would appear as a series of spectral lines whose
frequencies would be Doppler shifting continuously on both the spin period
of the brodcasting planet (a day for the Earth) as well as its orbital
period around the parent star (a year for the Earth). The planet-star
system resembles a spectroscopic binary in which one of the objects (the
planet) is a test particle in the gravitational field of a companion object
(the star) whose mass can be inferred from its emission spectrum through
follow-up observations with optical-infrared telescopes.  Under these
conditions, measurements of the planet's Doppler effect over a full orbit
would determine the inclination, eccentricity and semi-major axis of the
planet's motion around the star \footnote{The observed period and projected
velocity can be used to infer the inclination if the mass of the star is
known; see S. Shapiro \& S.A. Teukolsky, Black Holes, White Dwarfs, and
Neutron Stars: The Physics of Compact Objects, Wiley-New York (1983),
p. 254. The orbital period, projected velocity, and inclination, can then
be used to derive the semi-major axis, from which the eccentricity can be
deduced based on the period. The results can be tested for consistency by
observing the centroid shift of the radio image as a function of time, as
this would provide an independent measure of the distance and hence the
luminosity of the star.}.  The inferred orbital parameters can then be used
to estimate the temperature on the surface of the planet and to assess
whether it is likely to support liquid water or life as we know it.

The optimization of the detection algorithm could also lead to changes in
hardware or observing strategy with the goal of improving the prospects for
the detection of a SETI signal.  Although MWA-LFD provided the benchmark
example in this paper, our conclusions extend to other experiments such as
LOFAR, PAST, or the SKA.  In parallel to developing the detection and
analysis filter for these experiments, it would be prudent to compile an
extended catalog of all nearby stars with known planets where low-frequency
radio emission might be detectable.  This work will extend previous
analysis \cite{Turnbull} to include additional stars where planets were
found recently, as well as to span larger distances than those probed by
other SETI experiments.

\bigskip
\paragraph*{Acknowledgments}

We acknowledge support from the FQXi grant \#RFP1-06-22 and Harvard
university funds for A.L.; M.~Z.~is supported by the Packard and Sloan
foundations, NSF AST-0506556 and NASA NNG05GG84G. We thank the anonymous
referee for useful comments.

\bigskip
\bigskip
\bigskip
\bigskip
\bigskip
\bigskip
\bigskip
\bigskip
\bigskip
\bigskip
\bigskip
\bigskip

\begin{table}[htdp]
\caption{Characteristics of the largest components of radio emission from
the Earth (adopted from \cite{sullivan} and public web sites on {\it
BMEWS}). The power per solid angle $P_\Omega$ along the beam of military
radars ({\it BMEWS}) is larger by two orders of magnitude than the total
power entry in the table.}
\begin{center}
\begin{tabular}{||c|c|c|c|c|c|c||}
\hline\hline
Service & Freq.  & Transmitters & Max. Power  &  Bandwidth  & Power & Power/Hz\\ 
 & (MHz) &(No.) & per Tr. (W) & (Hz) & (W) & (W/Hz) \\ \hline
 & & & & & & \\
Military & $\sim 400$ & 10 & $2 \times 10^{8}$ & $10^{3}$ & $2 \times 10^{9}$ & $2 \times 10^6$\\
TV & 40-850 & 2000 & $5 \times 10^{5}$ & 0.1 & $10^{9}$ & $10^{10}$\\
FM & 88-108 & 9000 & $4 \times 10^{3}$ & 0.1 & $4 \times 10^{7}$ & $4 \times 10^{8}$  \\
\hline \hline
\end{tabular}
\end{center}
\label{earth} 
\end{table}%

\end{document}